\documentstyle[12pt]{article}
\input epsf.sty
\newdimen\psfigsize
\def\psfigure#1 #2 #3 #4 #5{
    \begin{figure}[tbhp]
    \vbox{
    \null\hskip#2\epsfxsize=#1 \epsfbox[0 0 4096 4096]{#4}
    \vskip 10truept
    \caption {#5 \label{#3}}
    \vskip 0.1truein plus0.2truein}
    \end{figure}
}
\def\pspagefigure#1 #2 #3 #4 #5{
    \begin{figure}[p]
    \vbox{
    \null\hskip#2\epsfxsize=#1 \epsfbox[0 0 4096 4096]{#4}
    \vskip 10truept
    \caption {#5 \label{#3}}
    \vskip 0.1truein plus0.2truein}
    \end{figure}
}
\def\psoddfigure#1 #2 #3 #4 #5 #6{
    \begin{figure}[tbhp]
    \vbox{
    \null\hskip#3\epsfxsize=#1 \epsfbox[0 0 4096 4096]{#5}
    \vskip -#1 \vskip #2 \vskip 10truept
    \vskip 10truept
    \caption {#6 \label{#4}}
    \vskip 0.1truein plus0.2truein}
    \end{figure}
}
\def\figurespace#1 #2 #3 #4 {
    \begin{figure}[tbhp]
    \vbox{
    \psfigsize=#1truein
    \vskip \psfigsize
    \vskip 10truept
    \caption {#4 \label{#3}}
    \vskip 0.1truein plus0.2truein}
    \end{figure}
}
\def\gnufigure#1 #2 #3 #4 #5 #6{
    \begin{figure}[tbhp]
    \vbox{
    \null\hskip#3\epsfxsize=#1 \epsfbox{#5}
    \vskip -#1 \vskip #2 \vskip 10truept
    \vskip 10truept
    \hbox{\null\hskip 1.0in \parbox[t]{4.5in}{ \caption {#6 \label{#4}} } }
    \vskip 0.1truein plus0.2truein}
    \end{figure}
}

\def\pbp{\bar\psi\psi}

\newcommand{\Plaq}{\Box}

\def\LP{\left(}		
\def\RP{\right)}	
\def\PAR#1#2{ {{\partial #1}\over{\partial #2}} }

\def\BE{\begin{equation}}
\def\EE{\end{equation}}
\def\BEA{\begin{eqnarray}}
\def\EEA{\end{eqnarray}}
\def\EL{\nonumber\\}
  
\newcommand{\la}[1]{\label{#1}}

\newcommand{\atantwo}{{\rm atan2}}
 
\begin{document}

\begin{titlepage}
\baselineskip=16pt
\rightline{\bf hep-lat/9607084}
\rightline{AZPH-TH/96-13}
\baselineskip=20pt plus 1pt
\vskip 1.5cm

\centerline{\Large \bf  Scaling functions for O(4) in three dimensions}
\vskip 1.5cm
\bigskip
\centerline{\bf Doug Toussaint}
\centerline{\it Physics Department, University of Arizona, Tucson, AZ
85721, USA}
\centerline{\sf doug@physics.arizona.edu}

\narrower
Monte Carlo simulation using a cluster algorithm
is used to compute the scaling part of the
free energy for a three dimensional O(4) spin model.  The results are
relevant for analysis of lattice studies of high temperature QCD.
\end{titlepage}

\vskip0.25in\centerline{Introduction}

The high temperature phase transition for QCD with two flavors of light
quarks is expected to be driven by chiral symmetry restoration, with an
order parameter having O(4) symmetry in the continuum
limit\cite{PisarskiandWilczek,RajagopalandWilczek,Rajagopal}.
Thus, near the transition we expect the scaling properties of a three
dimensional O(4) spin model.
For quark mass or temperature not too close to the transition, the
system would be expected to behave like mean field theory.  Recently Kogut
and Kocic have suggested that mean field behavior might describe the
system arbitrarily close to the critical point\cite{KogutandKocic}.
Finally, with Kogut-Susskind quarks on a nonzero lattice spacing, the
exact chiral symmetry is only O(2), and it is possible that lattice
simulations are better described by O(2) critical behavior.
In addition to its intrinsic interest as an indicator of the physics of
the transition, the form of the free energy near the critical point is
important in extrapolating the QCD equation of state from the quark
masses where lattice simulations are practical to the light quark masses
of the real world\cite{milc_eos96,milc_nt12}.

Assuming a second order transition\cite{UKAWA}, we expect the singular behavior of
thermodynamic observables near the transition to be universal, meaning
that the symmetry group of the order parameter and the dimension of the
system determine the critical exponents and the form of the singular part of
the free energy, up to normalization of the scaling variables.
(See, for example, \cite{AMIT})  The
critical exponents for O(4) and O(2) are well
known\cite{Kanaya_and_Kaya,Baker_Meiron_and_Nickel,ButeraandComi},
but the form
of the free energy, or the ``scaling function'', is only
poorly known.  An epsilon expansion result is
available\cite{epsilon_expansion},
quoted in Ref.~\cite{RajagopalandWilczek}.
Similarly, Monte Carlo calculations of critical exponents have been
used to study the critical behavior of high temperature
QCD\cite{Karsch_and_Laermann}, but to date
the full power of the scaling ansatz, namely comparison with the
universal scaling functions as well as critical
exponents, has not been brought to bear.

Here we use Monte Carlo simulation to compute an approximate scaling
function for O(4), to be used in comparing to Monte Carlo simulations of
QCD.

For the O(N) spin model we use the partition function:
\BE Z = \int [d \vec s] \exp\left( J\sum_{ij}\vec s_i \cdot \vec s_j + H
\sum_i s_{0i} \right) \ \ \ ,\EE
where $ij$ are nearest neighbor pairs on a (hyper)cubic lattice
in $d$ dimensions and $s_{0i}$ is the zero
component of $\vec s_i$.  Then the energy 
and magnetization are:
\BEA \langle E \rangle = \frac{1}{dV} \PAR{\log(Z)}{J} \EL
     \langle M \rangle = \frac{1}{V} \PAR{\log(Z)}{H} \ \ \ .\EEA
In QCD, using the normalization where the plaquette ($\Plaq$)
is three when all links are unity, the analogous equations are
\BEA \langle \Plaq \rangle = \frac{1}{2Vn_t} \PAR{\log(Z)}{6/g^2} \EL
     \langle \pbp \rangle = \frac{1}{Vn_t} \PAR{\log(Z)}{am_q} \EEA

\vskip0.25in\centerline{Parameterizing the scaling functions}

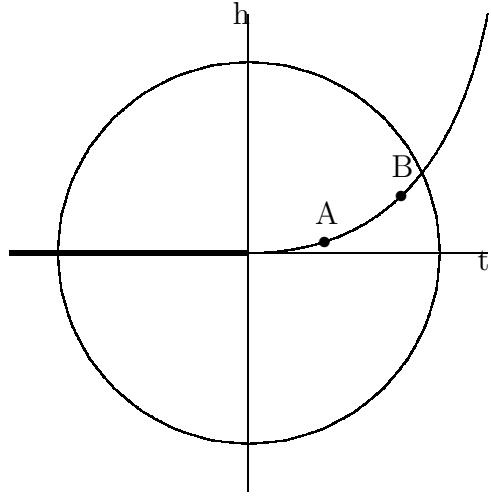
\begin{figure}
\setlength{\unitlength}{1.0in}
\begin{picture}(6.0,3.0)(-3.0,-1.5)  

\put(-1.25,0){\line(1,0){2.5}}
\put(0,-1.25){\line(0,1){2.5}}
\put(1.2,-0.09){t}
\put(-0.08,1.2){h}
\thicklines\put(-1.25, 0.000){\line(1,0){1.25}}
           \put(-1.25, 0.005){\line(1,0){1.25}}
           \put(-1.25,-0.005){\line(1,0){1.25}}\thinlines

\qbezier(1.00000,0.00000) (0.995185,0.0980171) (0.980785,0.19509)
\qbezier(0.980785,0.19509) (0.95694,0.290285) (0.92388,0.382683)
\qbezier(0.92388,0.382683) (0.881921,0.471397) (0.83147,0.55557)
\qbezier(0.83147,0.55557) (0.77301,0.634393) (0.707107,0.707107)
\qbezier(0.707107,0.707107) (0.634393,0.77301) (0.55557,0.83147)
\qbezier(0.55557,0.83147) (0.471397,0.881921) (0.382683,0.92388)
\qbezier(0.382683,0.92388) (0.290285,0.95694) (0.19509,0.980785)
\qbezier(0.19509,0.980785) (0.0980172,0.995185) (0,1)
\qbezier(0,1) (-0.0980171,0.995185) (-0.19509,0.980785)
\qbezier(-0.19509,0.980785) (-0.290285,0.95694) (-0.382683,0.92388)
\qbezier(-0.382683,0.92388) (-0.471397,0.881921) (-0.55557,0.83147)
\qbezier(-0.55557,0.83147) (-0.634393,0.77301) (-0.707107,0.707107)
\qbezier(-0.707107,0.707107) (-0.77301,0.634393) (-0.83147,0.55557)
\qbezier(-0.83147,0.55557) (-0.881921,0.471397) (-0.92388,0.382683)
\qbezier(-0.92388,0.382683) (-0.95694,0.290285) (-0.980785,0.19509)
\qbezier(-0.980785,0.19509) (-0.995185,0.0980172) (-1,0)
\qbezier(-1,0) (-0.995185,-0.0980171) (-0.980785,-0.19509)
\qbezier(-0.980785,-0.19509) (-0.95694,-0.290285) (-0.92388,-0.382683)
\qbezier(-0.92388,-0.382683) (-0.881921,-0.471397) (-0.83147,-0.55557)
\qbezier(-0.83147,-0.55557) (-0.77301,-0.634393) (-0.707107,-0.707107)
\qbezier(-0.707107,-0.707107) (-0.634393,-0.77301) (-0.55557,-0.83147)
\qbezier(-0.55557,-0.83147) (-0.471397,-0.881921) (-0.382684,-0.92388)
\qbezier(-0.382684,-0.92388) (-0.290285,-0.95694) (-0.19509,-0.980785)
\qbezier(-0.19509,-0.980785) (-0.0980172,-0.995185) (0,-1)
\qbezier(0,-1) (0.0980171,-0.995185) (0.19509,-0.980785)
\qbezier(0.19509,-0.980785) (0.290285,-0.95694) (0.382683,-0.92388)
\qbezier(0.382683,-0.92388) (0.471397,-0.881921) (0.55557,-0.83147)
\qbezier(0.55557,-0.83147) (0.634393,-0.773011) (0.707107,-0.707107)
\qbezier(0.707107,-0.707107) (0.77301,-0.634393) (0.83147,-0.55557)
\qbezier(0.83147,-0.55557) (0.881921,-0.471397) (0.923879,-0.382684)
\qbezier(0.923879,-0.382684) (0.95694,-0.290285) (0.980785,-0.19509)
\qbezier(0.980785,-0.19509) (0.995185,-0.0980172) (1.00000,0.0000000)

\qbezier(0,0)(1.0,0),(1.25,1.25)

\put(0.4,0.06){\circle*{0.05}} \put(0.35,0.15){A}
\put(0.8,0.30){\circle*{0.05}} \put(0.75,0.40){B}

\end{picture}
\caption{
A length rescaling by a factor of $b$ is accomplished by
changing the couplings $(t,h)$ at point A to $(b^{y_t}t, b^{y_h}h)$ at
point B by moving along the renormalization group trajectory (curved
line).  The trajectories may be labelled by their intersections with
the unit circle, so specifying the free energy on the unit circle,
together with the scaling ansatz which tells how the free energy
changes along a trajectory, specifies the free energy everywhere.
The discontinuity in the order parameter at $t<0$ and $h=0$ (heavy line)
implies that the derivative of the free energy is discontinuous there.
}
\label{RENFLOW}
\end{figure}

From invariance under a length rescaling by a factor $b$,
the critical part of the free energy should have the property:
\BE\la{scaling_eq} f_s(t,h) = b^{-d} f_s(b^{y_t}t,b^{y_h}h) \ \ .\EE
Here $t$ and $h$ are the scaling variables, with the critical point at
$(t,h)=(0,0)$, and $y_t$ and $y_h$ are the corresponding critical
exponents.  Other exponents can be expressed in  terms of $y_t$ and
$y_h$.
$t = (T-T_c)/T_0$ and $h=H/H_0$ are conventionally normalized by requiring that
$M(t=0,h)=h^{1/\delta}$ and $M(t<0,h=0)=(-t)^\beta$.
The free energy also has a nonsingular part.

The scaling ansatz, Eq.~\ref{scaling_eq}, implies that the magnetization
near the critical point
is determined by a universal scaling function, conventionally written as:
\BE\la{f_scaling_eq} \frac{M}{h^{1/\delta}} = f(t/h^{1/\beta\delta}) = f(x) \
\ .\EE
The normalization conditions on $t$ and $h$ then require that $f(0)=1$
and $f(x) \rightarrow (-x)^\beta$ as $x \rightarrow -\infty$.

In computing the energy and pressure of QCD, we require the plaquette,
analogous to the energy in the spin model, extrapolated to zero quark
mass, which is analogous to zero magnetic field.  The
magnetization (or $\pbp$ in QCD) is
$\frac{1}{V}\PAR{\log(Z)}{H} = \frac{1}{V H_0} \PAR{\log(Z)}{h}$
while the energy (or plaquette in QCD) is
$\frac{1}{V} \PAR{\log(Z)}{T} = \frac{1}{V T_0}\PAR{\log(Z)}{t}$.
Since the energy and magnetization are derivatives of the free energy
with respect to $t$ and $h$ respectively, information about one quantity
constrains the other.  In particular, the behavior of $\pbp$ in QCD
can help in extrapolating the plaquette to zero or small quark mass.
(It is important to enforce consistency of the plaquette and $\pbp$
in computing the equation of state.  Using a plaquette and $\pbp$ which
are not derivatives of the same free energy could lead to inconsistent
thermodynamics.)  Because our analysis requires that we handle the
energy and magnetization on the same footing, in addition to
Eq.~\ref{f_scaling_eq}, we develop a formulation of the
scaling ansatz which treats the magnetization and energy equally.

The scaling ansatz, Eq.~\ref{scaling_eq},
tells us that if we specify the singular free energy on
any circle in the $t,h$ plane, we have specified it for all $t,h$.
(See Fig.~\ref{RENFLOW})
In particular, if $g(\theta)$ is the scaling free energy on the unit circle in
the $t,h$ plane, then the rescaling factor $b$ which takes $t,h$ to the
unit circle is determined by
\BE\la{unit_eq} \LP b^{y_t} t \RP^2 + \LP b^{y_h} h \RP^2 = 1 \EE
For $y_t$ and $y_h$ positive this
clearly has a unique solution for $b>0$ given $t$ and $h$.
Although in general this cannot be solved analytically for $b$, it is
straightforward to do differentiations implicitly and solve the equation
numerically.
Then the singular free energy
(actually minus one times temperature times free energy per volume) is
\BE\la{g_scaling_eq} \frac{1}{V} \log(Z_s(t,h)) = b(t,h)^{-d}g(\theta(t,h)) \EE
where
$\theta(t,h) = \atantwo(b^{y_h} h, b^{y_t} t)$, and $g(\theta)$
is a universal function.

Equivalently, the relation between $t,h$ and $b,\theta$ can be expressed
as:
\BEA\la{formtwo}
&&b^{y_h}h = \sin(\theta) \EL
&&b^{y_t}t = \cos(\theta) \ \ \ .\EEA

After some differentiations, the magnetization and energy
can be expressed in terms of $g(\theta)$:
\BE M = \frac{1}{H_0} \LP \frac{-d}{b^{d+1}} \PAR{b}{h} g(\theta) +
b^{-d} g^\prime (\theta) \PAR{\theta}{h} \RP \EE
\BE E  =  \frac{1}{d T_0} \LP \frac{-d}{b^{d+1}} \PAR{b}{t}
g(\theta) + b^{-d} g^\prime(\theta) \PAR{\theta}{t} \RP \ \ \ ,\EE
where
\BEA
\PAR{b}{h}\big|_t &&= 
   \frac{-hb^{2y_h}}{y_t t^2 b^{2y_t-1}+ y_h h^2 b^{2y_h-1} } 
    = \frac{-\sin(\theta) b^{y_h+1}}{y_t \cos^2(\theta)+ y_h \sin^2(\theta)}\EL
\PAR{b}{t}\big|_h &&= 
   \frac{-tb^{2y_t}}{y_t t^2 b^{2y_t-1}+ y_h h^2 b^{2y_h-1} } 
    = \frac{-\cos(\theta) b^{y_t+1}}{y_t \cos^2(\theta)+ y_h \sin^2(\theta)}\EL
\PAR{\theta}{h}\big|_t &&= 
   t b^{y_t+y_h} + ht\PAR{b}{h} \LP y_h-y_t \RP b^{y_h+y_t-1} 
    = \frac{ y_t b^{y_h}\cos(\theta)}{y_t \cos^2(\theta)+ y_h\sin^2(\theta)}\EL
\PAR{\theta}{t}\big|_h &&= 
   -h b^{y_t+y_h} + ht\PAR{b}{t} \LP y_h-y_t \RP b^{y_h+y_t-1} 
    = \frac{-y_h b^{y_t}\sin(\theta)}{y_t \cos^2(\theta)+ y_h\sin^2(\theta)}\EL
\ \ \ .\EEA

Physical insight into the form of $g(\theta)$ comes from considering
special cases.

First, for $t>0$ and $h$ small:
\BEA
&&M = \frac{1}{H_0} g^\prime(0) t^{d/y_t} t^{-y_h/y_t}  \EL
&&E = \frac{ g(0)}{ T_0 y_t} t^{ d/y_t-1 }
\ \ \ .\EEA
Since $M$ must vanish here, we require $g^\prime(0)=0$.  In fact, we
expect the free energy to be an even function of $h$, with a cusp at
$h=0$ and $t<0$ due to the discontinuity of $M$ on this line.
One more differentiation of the energy will give the specific
heat $ C \approx t^{d/y_t-2} = t^{-\alpha}$.
We also find that the susceptibility is 
\BE \chi = \frac{1}{H_0^2}\LP \frac{d g(0)}{y_t}+g^{\prime\prime}(0)\RP
t^{(d-2y_h)/y_t} \ \ \ .\EE

For $t<0$ and $h$ small and positive,
\BEA
&&M = \frac{-1}{H_0} g^\prime(\pi) (-t)^{(d-y_h)/y_t} \approx
(-t)^\beta  \EL
&&E = \frac{- g(\pi)}{ T_0 y_t} (-t)^{ d/y_t-1 }
\ \ \ .\EEA

For $t=0$ and $h>0$,
\BEA
&&M = \frac{d}{H_0 y_h} g(\pi/2) h^{d/y_h-1} \EL
&&E = \frac{-1}{d T_0} g^\prime(\pi/2) h^{(d-y_t)/y_h}
\ \ \ .\EEA

From these expressions we get the following intuition about the scaling
free energy $g(\theta)$:
\begin{enumerate}
\item{ $g(0)$ controls the singular part of the energy for
$T>T_c$.}
\item{ $g(\pi$) controls the singular part of the energy for
$T<T_c$.}
\item{$g^\prime(\pi/2)$ controls the energy for $t=0$, $h \ne 0$.}
\item{$\lim_{\theta \uparrow \pi} g^\prime(\theta) = 
  -\lim_{\theta\downarrow -\pi} g^\prime(\theta)$
controls the expectation value of $M$ for $T<T_c$.}
\item{$g(\pi/2)$ controls $M$ for $t=0$ and $h \ne 0$.}
\end{enumerate}
Here it is convenient to choose $T_0$ and $H_0$ so that
$H_0 M(t=0,h) = h^{1/\delta}$ and $H_0 M(t<0,h=0) = (-t)^\beta$.
The normalization conditions on $t$ and $h$ then require that $g(\pi/2)
= y_h/d$ and $g^\prime(\pi) = -1$.
When it is necessary to distinguish, we will call $H_0$ for the ``$f(x)$"
and ``$g(\theta)$'' forms $H_f$ and $H_g$ respectively.  Similarly we
distinguish $T_f$ and $T_g$.  They are related by $H_f=H_g^{\delta+1}$
and $T_f=T_g H_g^{1/\beta}$.

\vskip0.25in\centerline{Simulations}

Monte Carlo simulations were run on $16^3$, $24^3$, $32^3$, $40^3$,
$48^3$ and
$64^3$ lattices using a multiple cluster updating
algorithm\cite{WOLFF}.

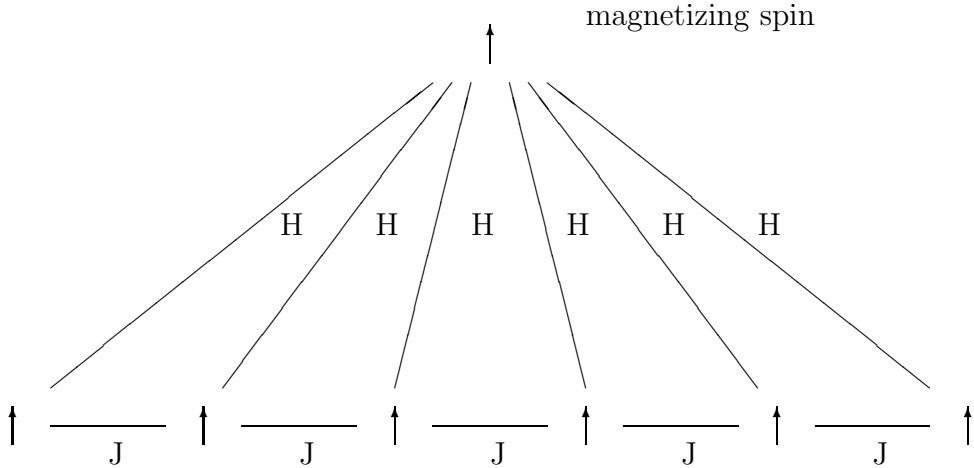
\begin{figure}
\setlength{\unitlength}{1.0in}
\begin{picture}(6.0,3.0)(-3.0,0.0)  

\put(-2.5,0.4){\vector(0,1){0.2}}
\put(-1.5,0.4){\vector(0,1){0.2}}
\put(-0.5,0.4){\vector(0,1){0.2}}
\put( 0.5,0.4){\vector(0,1){0.2}}
\put( 1.5,0.4){\vector(0,1){0.2}}
\put( 2.5,0.4){\vector(0,1){0.2}}
\put( 0.0,2.4){\vector(0,1){0.2}}
\put( 0.5,2.6){magnetizing spin}

\multiput(-2.3,0.5)(1.0,0.0){5}{\line(1,0){0.6}}
\multiput(-2.0,0.3)(1.0,0.0){5}{J}
\put(-2.3,0.7){\line( 5,4){2.0}}
\put(-1.4,0.7){\line( 3,4){1.2}}
\put(-0.5,0.7){\line( 1,4){0.4}}
\put( 0.5,0.7){\line(-1,4){0.4}}
\put( 1.4,0.7){\line(-3,4){1.2}}
\put( 2.3,0.7){\line(-5,4){2.0}}
\multiput(-1.1,1.5)(0.5,0){6}{H}

\end{picture}
\caption{
Cluster updating with a magnetic field.  A single ``magnetizing spin''
is coupled to every spin in the lattice by a bond with strength $H$.
}
\label{UPDATE}
\end{figure}

To use a cluster updating algorithm with a nonzero magnetic field,
just imagine that in addition to the regular bonds with strength $J$
connecting neighboring spins, each spin is connected to a fake
``magnetizing spin'' by a bond of strength $H$, as illustrated
in Fig.~\ref{UPDATE}.  Then break both ``J bonds'' and ``H bonds'' and
update clusters according to the usual cluster algorithm\cite{WOLFF}.
The ``magnetizing spin'' is a member of a cluster, and is reflected just
like any other spin.  When evaluating the magnetization of the lattice,
we take the components of the lattice spins in the current direction of
the magnetizing spin.

Results for the magnetization of the O(4) model are plotted in
Fig.~\ref{MAG_BEST}.  Results from the largest lattice size run at each
point are shown.  The remaining finite size effects are about the same
size as the statistical error bars.  Then, in Fig.~\ref{F_O4} the
results for $h=0.002$, $0.005$, $0.01$ and $0.02$ are plotted in the form in
Eq.~\ref{f_scaling_eq}.  Here we used the values for the critical
coupling and exponents from Kanaya and Kaya\cite{Kanaya_and_Kaya}.

Then we fit the magnetization results to find an approximate scaling
function $g(\theta)$ for the free energy in the form
in Eq.~\ref{g_scaling_eq}.  A simple parameterization of $g(\theta)$ which
satisfies the normalization conditions on $g(\pi/2)$ and $g^\prime(\pi)$
is
\BEA
\la{g_form_eq}
g(\theta) &&= y_h/3 + 2(\cos(\theta/2)-\sqrt{1/2}) \EL
&&+ a_0 \cos(\theta) \EL
&&+ a_1 ( cos(3\theta/2) + 3\cos(\theta/2)-2\sqrt{1/2} ) \EL
&&+ a_2 ( cos(2\theta) + 1 ) \EL
&&+ a_3 ( \cos(5\theta/2) - 5\cos(\theta/2) + 6\sqrt{1/2} ) \EL
\EEA
($d=3$ in this equation.)

In this fit I used the energy and magnetization for $0.89 < J < 0.99$
and $H=0.005$ and $0.002$.  The free energy also included an analytic
part $f_A = C_{H2} h^2 + C_{J1} t + C_{J2} t^2 + C_{J3} t^3$.
The resulting $g(\theta)$ is plotted in Fig.~\ref{g_theta_fig}.
$T_g$ and $H_g$ were 0.44 and 1.31 respectively.
The magnetization corresponding to this free energy is
also plotted in Fig.~\ref{mag_o4_3d}.
In principle, the critical exponents $y_t$ and $y_h$, and the critical
coupling are also parameters in this fit.  However, to get these parameters
to the same accuracy as has already been done by
Kanaya and Kaya\cite{Kanaya_and_Kaya} or
Butera and Comi\cite{ButeraandComi} would
require a careful correction for finite size effects, and care in
using only data for small enough $t$ and $h$ that corrections to scaling
are small.  Therefore, the exponents and critical coupling were fixed
to those found by Kanaya and Kaya.
Because of the remaining finite size effects and corrections to scaling,
the $\chi^2$ of this fit was very bad (243 for 30 degrees of freedom).
However, since the results
are already many times more accurate than the QCD data with which we
intend to compare, there is little incentive to make the necessary
corrections.
This scaling function was then converted to the ``$f(x)$'' form (by
computing the resulting magnetization as a function of $t$ for 
$h=0.002$ and plotting according to Eq.~\ref{f_scaling_eq}.), and
plotted as a solid line in Fig.~\ref{F_O4}, where it can be seen to
describe the magnetization quite well.
(It is necessary to convert the normalization of $t$ and $h$
used in the ``$g(\theta)$'' form to the conventional normalizations
for the ``$f(x)$'' form:  $H_{f} = H_{g}^{\delta+1}$
and $T_{f} = T_{g} H_{g}^{1/\beta}$.)
In this figure I have also included the mean field form of the scaling
function and the epsilon expansion form.

Fig.~\ref{F_O4} also shows the properly normalized asymptotic form
for $f(x)$ as $x \rightarrow -\infty$, $x^{-\beta}$.
It can be seen that the scaling
function approaches this asymptotic form quite slowly.  This is the
region where the long distance physics is dominated by the Goldstone
bosons.  In particular, we expect that for $t<0$ and $h$ small, 
the magnetization takes the form $M = M(t,0) + A h^{1/2}$, so that
the susceptibility diverges at $h=0$ for all $t<0$\cite{GOLDSTONE_SUSC}.
The fitting function Eq.~\ref{g_form_eq} should really be modified
to support this behavior at $\theta=\pi$, but this problem seems to
occur in a region beyond where these O(4) results,
and the QCD results to which they will be compared, are taken.

The Monte Carlo scaling function and the epsilon expansion are in good
agreement for $t<0$ because of the normalization condition on $t$,
$M(t<0,h=0)=(-t)^\beta$.  Had we chosen
the equally sensible normalization condition $\chi = \frac{dM}{dh} =
t^{-\gamma}$ for $t>0$ and $h=0$, we would have found agreement of
the Monte Carlo and epsilon expansion for $t>0$ with a discrepancy for $t<0$.


\psfigure 6.0in 0.0in {MAG_BEST} {mag_best.ps} {
Magnetization in the O(4) model for $h=0.002$, $0.005$, $0.01$, $0.02$ and $0.05$.
Octagons are for $L=16$, squares for $L=24$, bursts for $L=32$,
diamonds for $L=40$, decorated diamonds for $L=48$ and decorated
plusses for $L=64$.
}

\psoddfigure 7.0in 6.6in  -0.75in {F_O4} {f_o4.ps} {
O(4) magnetization for $h=0.002$, $0.005$ and $0.01$ plotted as a
scaling function in the conventional form.  I use the results of Kanaya
and Kaya\protect\cite{Kanaya_and_Kaya} for the critical coupling and critical exponents.
In this plot the points for $h=0.05$ are plotted with plusses, those for
$0.02$ with crosses, $0.01$ with diamonds, $0.005$ with octagons and
$0.002$ with squares.
Also shown are the asymptotic forms $f(x) \approx (-x)^\beta$ as $x \rightarrow
-\infty$ (a) and $f(x) \approx C x^{-\gamma}$ as $s \rightarrow \infty$
(b).
A four parameter fit to the scaling function from the
``$g(\theta)$'' form is shown with a solid line (c), running through the
entire graph.
The mean field scaling function, and the second order epsilon expansion
scaling function \protect\cite{epsilon_expansion} are shown,
labelled ``$\epsilon$'' and ``mf''.
}

\psfigure 5.0in 0.5in {g_theta_fig} {g_theta.ps} {
Four parameter fit to the scaling function $g(\theta)$ for the O(4) free energy.
}


\gnufigure 7.0in 7.0in -0.5in {mag_o4_3d} {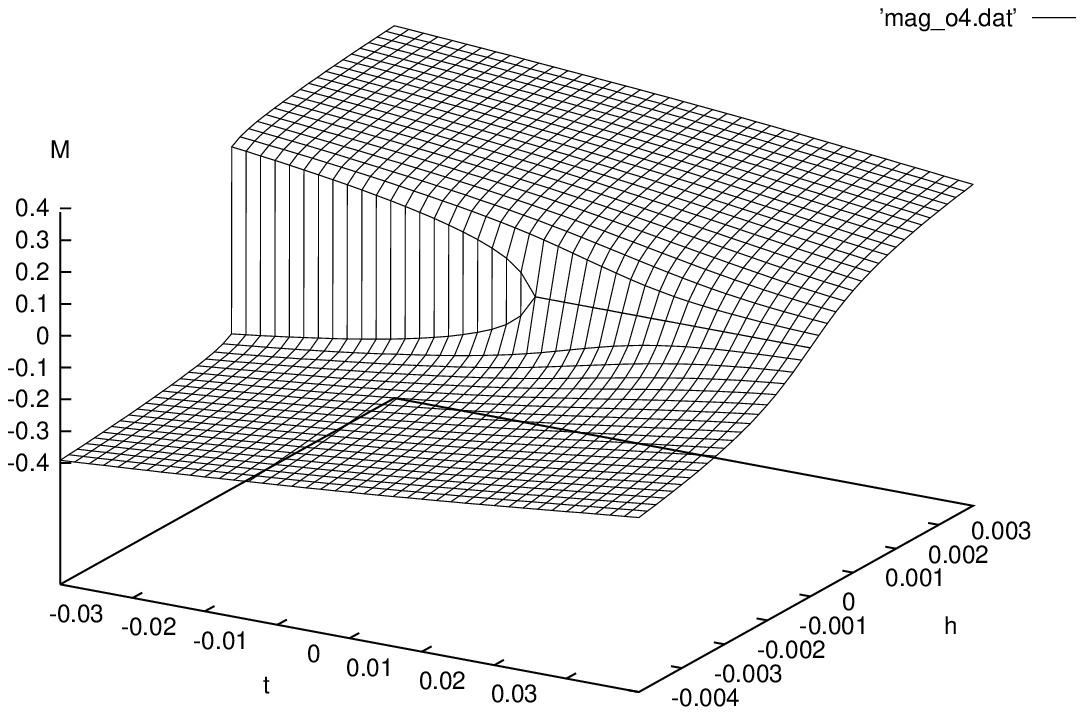} {
The scaling part of the magnetization for O(4), corresponding to the
scaling function in Fig.~\protect\ref{g_theta_fig}.
}

\vskip0.25in\centerline{Acknowledgements}

Tom Blum, Carleton DeTar and Bob Sugar have contributed greatly
to this work.
I thank Joe Rudnick for a valuable conversation, particularly for
pointing out to me the diverging susceptibility along the $-t$ axis.
This work was supported by DOE grant DE-FG03-95ER-40906.

\end{document}